# Petting Pen for Stress Awareness and Management in Children

Jing Li[1, *], Pinhao Wang[1], Emilia Barakova[1], Jun Hu[1]

1. Eindhoven University of Technology, Eindhoven, The Netherlands

We found that children in elementary school often experience stress during task performance. Limited coping skills and lack of stress awareness restrict children's ability to manage their stress. Many designs and studies have proposed different stress detection and intervention solutions. Still, they often overlook the potential of enhancing everyday objects and actively sensing stress-related behavioral data during human-product interaction. Therefore, we propose Petting pen as an interactive robotic object for children to manage their stress during task performance. It detects and validates stress and further intervenes in stress during a process of natural writing and relaxation interactions. The design is an iteration based on our previous research results of a stress-aware pen, enhanced with tactile needs, robotic interaction, and integration of behavioral and bio-sensing capabilities. Petting pen is supposed to bridge the gap between robots and everyday objects in mental health applications for children.

## Introduction

Stress from study and heavy workload have significantly contributed to mental health problems in children from East Asian countries such as China, Japan and Korea [28]. The pressures associated with homework and tasks are one of the primary sources of stress in elementary school children from these countries. Excessive and long-term stress from studying may lead to severe psychological problems such as depression, anxiety and even suicidal thoughts [17]. Unlike adults, who are more capable of being aware of self-stress states, children, especially at the elementary school level, cannot recognize and cope with the day-to-day stress that confronts them during study and task performance [11]. Research shows that most children do not receive support or treatment for mental health difficulties [13]. Due to the vulnerability of children in stressful situations, it is essential to consider alternative ways and methods during the study process to promote mental well-being.

Real-time stress detection and intervention technologies play an essential role in monitoring and improving mental health in children [4]. Many smart wearable devices [29] have been developed to monitor children's mental health through physiological data. These devices often do not embody the potential for behavior change. And most of the devices provide limited contextual information and overlook stress-related behavioral data.

Our previous work [15] has shown the potential relationship between stress and handwriting behaviors in children during task performance. These stress-related behaviors can be captured during human-product interactions in children's daily life. Thus, to help young children manage their stress during study and task performance unobtrusively, we propose using an everyday object already used in children's daily life but enhanced with robotic technologies.

We explored which types of interactions with everyday objects were potential stress indicators for elementary school children in the context of studying and task performance at home. Petting pen is therefore designed for monitoring stress and providing relaxing interaction with children. It detects and responds to stress-related behaviors and physiological signals of children. As an everyday robotic object, Petting pen helps children be aware of their stress states through tangible interaction.

Yet, Petting pen is still a conceptual design which was iterated based on the design and research results of Apen [15], which has been prototyped and tested in children. We enhanced Petting pen with robotic or animal-like behaviors to indicate stress states and interact with children during task performance at home.

## Related Work

### Stress Measurement

Stress is categorized into three types: acute, episodic, and chronic stress [2]. Acute stress and episodic acute stress last for a short period, making people anxious and frightened [1]. Chronic stress lasts longer than the other two types of stress, and can occur as long-standing pressure or accumulation of a lot of short-term stressful events [6].

Stress also can be classified into physiological stress and perceived stress. Physiological stress can be detected mainly in an individual's physiological response, and perceived stress is based on an individual's mental appraisal and psychological response [26]. Physiological stress can be quantified from several biomarkers, such as heart rate (HR), heart rate variability (HRV), skin conductance (EDA), and blood pressure (BP) [24].

Perceived stress is often measured with a clinical approach, such as a periodical self-report collected from the individuals [26]. Psychological stress has a more collective context from the perspective of data collection, but it is highly subjective, resulting in low availability and low credibility [3]. Nevertheless, researchers believe there is a coherency between psychological and physiological stress [6].

People respond to stress in different ways, physiologically, psychologically, and behaviorally[5]. Significantly, children appear to exhibit a clear behavioral response under stress. The anxious behaviors can be observed in object manipulation (e.g. playing with an object) [5, 9]. Therefore, embedding sensors in these related objects could help detect stress-related behaviors, providing information on stress levels.

### Smart and Interactive Everyday Objects

There is a trend that more and more self-tracking technologies are embedded in everyday objects [7]. These everyday objects can be integrated into a system that can actively or passively sense the user behavior and interpret it into stress-related information [5].

Several common requirements exist for designing stress detection devices for children with ASD [14]. It has to be non-invasive[1]; it should be easy to use and easy to learn [23]; and the cost of purchasing and maintaining the device should be affordable for families [27].

Everyday objects have a high potential for collecting contextual information in daily life. When everyday objects are enhanced with emerging technologies like sensors, actuators and wireless connections, collected data can help develop an understanding of the minutiae of one's daily life [21]. Also, everyday objects can be useful, informative, and connected for data collection in an unobtrusive way [25].

Those pervasive everyday objects can be enhanced and used to collect stress-related information through human-product interactions [16, 18, 19] to help recognize stressful situations and identify stressful behavior patterns [20]. There is potential for everyday objects to be enhanced and even integrated into a connected stress detection system that is able to actively and passively sense behavioral input from the user's interaction, and the system can interpret behavioral information into stress-related information.

A lot of design research has tried to detect stress through behavioral data such as movements or activities [10, 20], and physiological data such as HRV [12, 22] and EDA [8], which were collected from sensors embedded in objects.

## Design Process

Petting pen as the design iteration based on the user testing results of Apen [15] focuses on improving the user interaction with the pen. A pen as an everyday object was identified as being associated with stress-related behaviors in children during studying and task performance through interviews and observations. We made the first prototype, named Apen, as Figure 1 shows, which embedded pressure sensors, LED lights and vibration motors for the interaction. We tested Apen with four neurotypical children while performing their tasks. The results revealed that 1) There is an apparent correlation between physiological stress, perceived stress and writing behaviors from the collected data. 2) Lights and vibration as stress indications did not significantly increase the awareness of stress in children or change children's behaviors. 3) The means of interaction are limited and hard for children to understand.

In order to further develop the pen, we first conducted user interviews and user questionnaires with 92 children in elementary school. We found that most children have a limited understanding of what stress is. They tend to associate stress with destructive emotion, low morale and alertness.

However, most children are interested in seeing the indication or concretization of their stress states during studying. Moreover, the most popular intervention for children to deal with high-stress situations during task performance is taking a break and interacting socially with friends, parents, or pets. Taking the findings into design consideration, the iteration should offer concrete stress indications that link to children's interpretation of stress. And the iteration should also provide a chance for children to relax or take a break while sensing the high-stress singles. Following this, we transferred findings in a brainstorming session to generate a more detailed design in the form of a pen. We converged design solutions by evaluating the level of interaction and automation, the usability, and the acceptance of children.

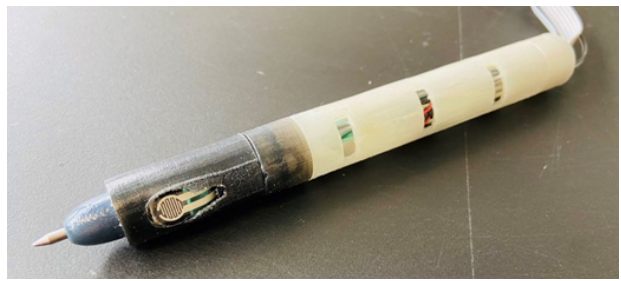

**Figure 1:** First prototype - Apen.

## Final Design

## Design Overview

As Figure 2 shows, Petting pen is designed for children to aid stress management during studying and task performance. It detects children's stress-related writing behaviors and physiological data and provides animal-like reactions as feedback. In this way, it aims to increase the stress awareness of children and intervene with stress through relaxing interactions. As an interactive robotic object, Petting pen is enhanced from an everyday object of children in the context of task performance. We enabled Petting pen with a biomimetic interface to engage children's interaction besides general writing purposes. Considering tactile needs and other relaxation strategies in stress intervention of children, we then finalized the design of Petting pen for stress management in children.

## Working Principle

Petting pen is able to detect handwriting and hand-holding behaviors when it is in use. A machine learning (ML) system supports the analysis and classification of stress-related behavioral data. When the potential stress behaviors are detected, Petting pen's main body (tail-looking) starts curving, as the left side of Figure 3 shows. The higher and longer stress-related behaviors are detected, the more deformation the tail becomes. The pen is not able to be used when the tail is entirely loppy. To reform the initial shape, the pen has to rest or react to the physiological data detected through the tail part.

The tail sways gently, as the middle in Figure 3 shows, if no physiological stress signal is detected, whereas the tail is puffed up and twitching, as the right side of Figure 3 shows, to alert the user that stress signals are detected.

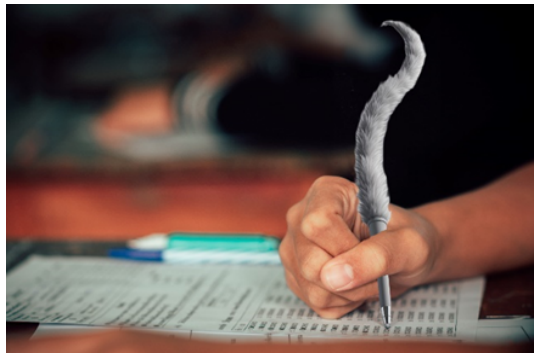

**Figure 2**: Petting pen in the context of studying.

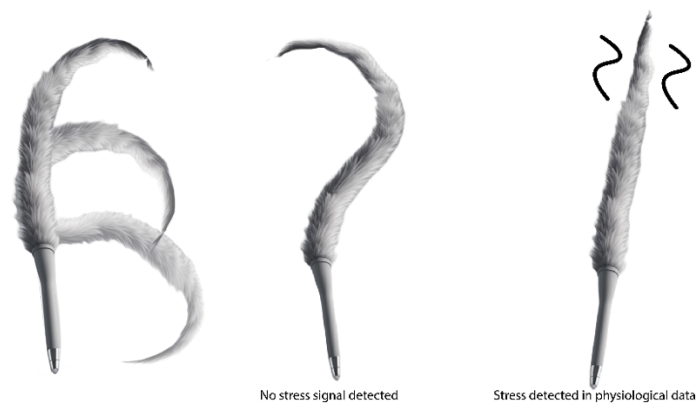

**Figure 3:** Curving tail in stress (The left side) & Different reactions on physiological data (The middle and right side).

## Interactive Experience

Figure 4 visualizes the main interactive stress management process between children and Petting pen. When children use the pen without stress, Petting pen has a waving tail occasionally as a relaxing behavior. While children keep studying and feeling stressed, they express stress on the writing and hand-holding pressure. Petting pen captures these stressful behaviors and starts curving and becoming lobby. Children may or may not notice the deformation of the pen, but they can no longer use the pen to write at some point since the deformation. At this moment, Children have to take a small break from their studying to pet their pen. The pen shows the hooked tail as an unsure behavior regarding children's mental state. The hooked tail guides children to pet and touch it while detecting their physiological data and validating if they are in a high-stress state. When the tail starts swaying or flicking softly, it indicates no physiological stress detected, and children can continue working with Petting pen. If there is prominent physiological stress input, the tail will puff up or straight up and quiver at the tip. In this way, children sense the tactile feedback as a stress alert and can further intervene in their stress by taking an extended break or other activities. In the high-stress state, Petting pen behaves challenging to be used with the purpose of providing a distraction or an unplanned break for children during their task performance.

## Technical Details

Figure 5 explains the embedment of the sensors and actuators in the pen. Petting pen is embedded with pressure and force sensors in the front body and the point where it detects hand-holding pressure and writing pressure. Our previous study[15] had preliminary results that revealed the correlation between hand-holding, writing pressure and stress level. With the machine learning technique, Petting pen can have a personalized stress behavior model that captures each individual's stress-related writing behavior data. There are optical sensors on the main body of the pen that detect contact and stress-related physiological data. According to data and the analyzed stress state, actuators, including connection joins, motors will collaborate and react differently.

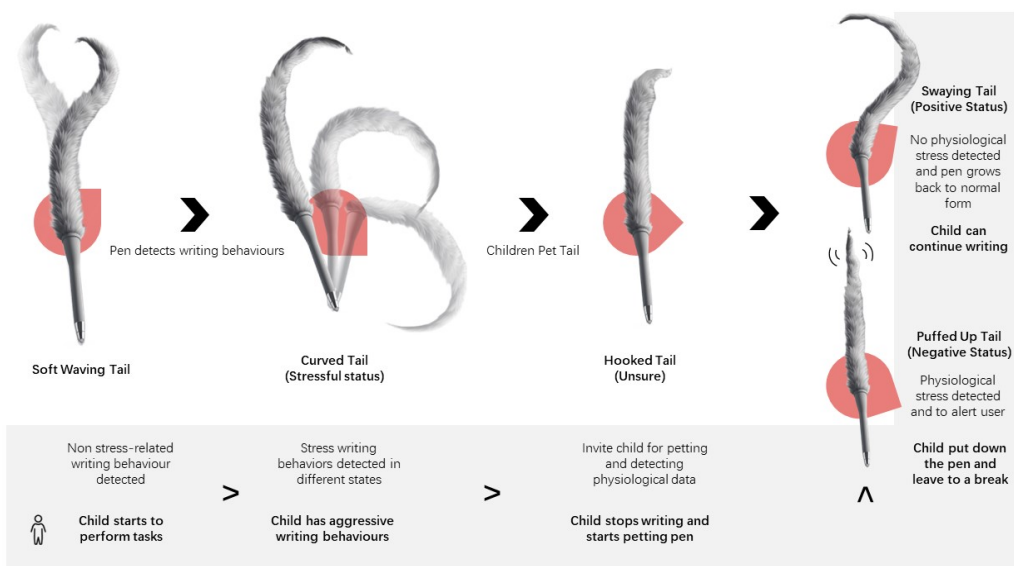

**Figure 4:** Child-Product Interaction map.

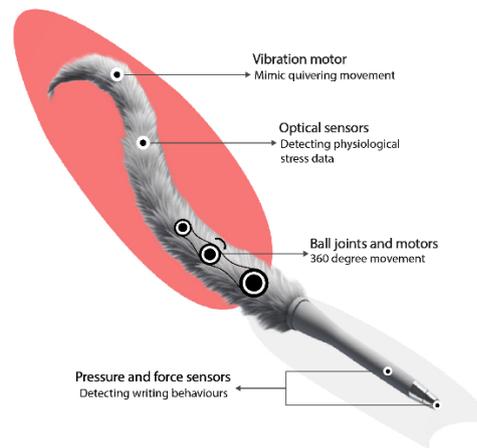

**Figure 5:** Sensors and actuators in Petting pen.

# Conclusion

Petting pen is an enhanced everyday robotic object supporting children in managing their stress during study and task performance. Based on the natural interaction of an everyday object in children's studying environment, we enhanced a pen with animal-like robotic behavior. It is enabled to detect stress-related writing behaviors and physiological stress data during the interaction. The design needs to be further prototyped and validated in the field study regarding the usability and effectiveness of improving stress awareness and relaxation. We also expect several iterations in the further design process to enhance the design of Petting pen. Furthermore, a personalized ML model also needs to be developed to study an individual's stress-related writing behaviors. We aim to connect Petting pen with other smart devices as a system of Internet of Things for supporting children's mental health in different contexts of everyday life.